\documentclass[twocolumn,showpacs,prbl,amsmath,amssymb]{revtex4}

\usepackage{graphicx}\usepackage{dcolumn}\usepackage{bm}

\begin{document}

\def\bk{{\bf k}}
\def\bq{{\bf q}}
\def\bp{{\bf p}}
\def\bQ{{\bf Q}}
\def\br{{\bf r}}
\def\bA{{\bf A}}
\def\ve{\varepsilon}

\title{The Hall conductivity in unconventional charge density wave systems}
\author{D. N. Aristov}
\altaffiliation[On leave from ]
{Petersburg Nuclear Physics Institute, Gatchina  188300, Russia.}
\author{R. Zeyher}
\affiliation{
Max-Planck-Institut f\"ur Festk\"orperforschung, Heisenbergstra\ss e 1, 
70569 Stuttgart,
Germany}

\date{\today}

\begin{abstract}     
Charge density waves with unconventional order parameters,
for instance, with d-wave symmetry (DDW), may be relevant in the 
underdoped regime of high-T$_c$ cuprates or other quasi-one or
two dimensional metals. A DDW state is characterized by two branches of 
low-lying electronic excitations. The resulting quantum mechanical current has 
an inter-branch component which leads to an additional mass term in the
expression for the Hall conductivity. This extra mass term is 
parametrically enhanced near the ``hot spots'' of fermionic dispersion
and is non-neglegible as is shown by numerical calculations of the Hall 
number in the DDW state.
\end{abstract}

 \maketitle

%%%%%%%%%%%%%%%%%%%%%%%%%%%%%%%%%%%%%%%%%%%%%%%%%%%%%%%%%%%%%%

%\section{introduction}

Recently, the interest in charge density waves with unconventional
order parameters has increased\cite{Maki,Neto,Cappelluti,Chakravarty}.
In particular, it has been shown that a charge density wave with d-wave
symmetry (DDW) represents a stable state of the $t-J$ model in the
large-N limit in certain doping and temperature regions\cite{Cappelluti}.
It thus may be intimately related to the pseudogap phase of 
high-T$_c$ superconductors\cite{Cappelluti,Chakravarty}. 
The presence of a DDW state should also cause
changes in transport coefficients\cite{WKim02}.
Refs.\cite{Chakravarty02,Norman03} discuss DDW-induced changes in the Hall effect 
on the basis of the standard formula\cite{Ziman}, which is applicable to
an usual metal. In the present paper we argue that a careful reconsideration
of the Hall coefficient for the case of a DDW state results in an additional 
term to the usual expression. This term enhances the change in
the Hall number due to onset of a DDW order parameter.

Qualitatively, the appearance of the new term can be understood as follows. It
is known from the band theory of metals that the quantum mechanical current
operator consists of two parts, the intra-band derivative with respect to 
the wave vector
and the term describing the interband transitions. Usually, the interband energy
spacing is large enough to neglect the influence of the interband current term.

In the DDW state the situation is different. The
charge density wave with momentum $\bf Q$ couples 
electrons which differ in momentum by $\bf Q$. Taking, for instance, 
a square lattice and  ${\bf Q} ={(\pi,\pi)}$ a two-band
picture is obtained in the reduced magnetic Brillouin zone. There exists
regions around certain wave vectors ${\bf K}$, the so-called ``hot-spots'',
where the quasi-particle energies of both bands are close to the Fermi level.
The influence of these hot spots determines the changes in the Hall
conductivity, as shown in Ref.\cite{Chakravarty02}. We find that the 
interband contribution to the current is particularly important in the vicinity
of the hot spots, leading to significant changes in the theoretical predictions.

From a broader viewpoint, the necessity of inclusion of the interband current
terms is known for the case of almost degenerate electron spectra \cite{LL9}
and for the case of the electromagnetic response in nodal (d-wave)
superconductors.  On a formal level, it can be illustrated as follows.
Near the point of degeneracy, ${\bf K}$, the spectrum can be represented as
$\epsilon_{{\bf k}+{\bf K}} \sim \sqrt{k_x^2 + k_y^2}$. The inclusion of the external vector
potential through the Peierls substitution, ${\bf k} \to {\bf k} - e{\bf A}$,
leads to a troublesome non-analyticity of the fermionic action on ${\bf A}$, 
i.e.  $\sqrt{|{\bf A}|^2}$. The recipe for the correct treatment of the 
electromagnetic response in such cases is well known\cite{LL9}. It amounts to  
retaining the non-diagonal form of the Hamiltonian, which is analytic in ${\bf 
A}$ and contains the interband currents, until the end of the calculation. 

%\section{basic formulas}

The mean-field Hamiltonian in the DDW state is

       \begin{eqnarray}
       {\cal H}&=& \sum_{{\bf k},\sigma}
       \left[
       \xi_{\bf k} a^\dagger_{{\bf k}\sigma} a_{{\bf k}\sigma}
       + i\Delta_{\bf k} a^\dagger_{{\bf k}\sigma} a_{{\bf k}+{\bf Q},\sigma}
        +h.c. \right]
       \label{H}
       \end{eqnarray}
Taking nearest and next-nearest neighbor hoppings $t$ and $t'$
into account and putting the lattice constant of the square lattice
to unity, the electronic
dispersion is $\xi_{\bf k}= -2t (\cos k_x +\cos k_y) +4t' \cos k_x
 \cos k_y -\mu$. In the following we also will use the abbreviations 
$ \xi_\pm = (\xi_{\bf k} \pm \xi_{{\bf k}+{\bf Q}})/2$.
The d-density wave the order parameter is of
the form $\Delta_{\bf k} = \Delta_0(\cos k_x - \cos k_y) = 
- \Delta_{{\bf k}+{\bf Q}}$.

In terms of two-component fermion operator
$\Psi^\dagger_{{\bf k}\sigma} = \left( a^\dagger_{{\bf k}\sigma},
 a^\dagger_{{\bf k}+{\bf Q},\sigma} \right) $
the Hamiltonian becomes
$       {\cal H}= \sum_{{\bf k},\sigma}
       \Psi^\dagger_{{\bf k}\sigma} {\hat H}_{\bk} \Psi_{{\bf k}\sigma} $
with
       \begin{equation}
       {\hat H}_{\bk} =
       \begin{pmatrix}
        \xi_{\bf k} &  i \Delta_{\bf k} \\
        -i \Delta_{\bf k} &\xi_{{\bf k}+{\bf Q}}
       \end{pmatrix}.
       \label{Ham-mat}
       \end{equation}
It can be diagonalized by the
unitary transformation $U = \exp i\sigma^1 \theta_{\bf k}$ with
         \[
         \theta_{\bf k} =  (1/2)\arctan (\Delta_{\bf k}/\xi_-),
         \]
where $\sigma^i$  denote the Pauli matrices.
We have  $U \hat H U^\dagger = diag
 (\varepsilon_{1},\varepsilon_{2})\equiv
\hat h$ and  the new quasiparticle energies are
       \begin{equation}
       \varepsilon_{1,2}= \xi_{+} \pm
       \left[ \xi_{-}^2 +\Delta_{\bf k}^2 \right]^{1/2}.
       \end{equation}
The fermionic Green's function  is given by
       \begin{eqnarray}
       {\hat G}_{{\bf k}\sigma}(i\omega) &=&
        \left(i\omega -\hat H \right)^{-1},
       \label{defmatrG}
       \end{eqnarray}
and it is diagonalized by the same matrix $U$. We write  ${\hat
 G}=U^\dagger \hat g
U$ with  $ \hat g = (\omega - \hat h)^{-1} $.
%\section{conductivity below the transition temperature}
The external vector potential is included into the Hamiltonian
by the Peierls substitution
${\hat H}_{\bk} \to {\hat H}_{\bk -e {\bf A} } .$

In what follows we use the  Kubo approach within the linear response
theory. It expresses the d.c.\  conductivity tensor,
$\sigma_{\alpha\beta}$, in terms of the current-current correlation
function, in the limit of static uniform external fields. \cite{Mahan}
In case of point-like impurity scattering, allowing the
neglectance of the vertex corrections to the corresponding diagrams,
the standard derivation leads to the  following formula
      \begin{equation}
      \sigma_{\alpha\beta} 
      = - e^2
      \int \frac{dx d\bk}
       {(2\pi)^3} \frac{ \partial n(x)}
      {\partial x}
      {\rm Tr}\left[
      {\hat V}_{\bk}^\alpha
      {\hat G}^A_{\bk}(x)
      {\hat V}_{\bk}^\beta
      {\hat G}^R_{\bk}(x)   \right],
      \label{conduc2}
      \end{equation}
where $n(x)=(e^{x/T}+1)^{-1}$ is the Fermi function, the summation over the spin
index has been performed, and the integration over $\bk$ 
refers to the magnetic Brillouin zone in order to avoid double counting.
Here ${\hat G}^{A(R)}$ denote the advanced (retarded) Green's
function defined on the real energy axis $x$. The
group velocity, corresponding to the microscopic quantum-mechanical
current, is
            \begin{equation}
	    {\hat V}_{\bk}^\alpha =
	    {\partial{\hat H}_{\bk}}/{\partial k_\alpha}
	    \equiv\partial ^\alpha {\hat H}_{\bk}.
	    \end{equation}
Using the above definitions the expression for the conductivity
can be  rewritten as $ {\rm Tr}\left[ {\hat V}^\alpha {\hat G}^A {\hat V}^\beta
{\hat G}^R \right] = {\rm Tr}\left[ {\hat v}^\alpha {\hat g}^A {\hat v}^\beta
{\hat g}^R \right]$ with $ {\hat v}^\alpha  = U^\dagger {\hat V}^\alpha U $ and
\[ {\hat v}^\alpha  = \partial^\alpha \hat h +[ (U^\dagger  \partial^\alpha
U), \hat h] \equiv {\cal D}^\alpha \hat h .\]
The last equation shows that in the basis which diagonalizes the
Hamiltonian, the current  operator  ${\hat v}^\alpha $ is
defined by  a "covariant"  derivative, ${\cal D}^\alpha$, which differs from the
usual derivative by the Christoffel symbol. Explicitly,
the current operator is given  by
            \begin{eqnarray}
          {\hat v}^\alpha & =&
	  \begin{pmatrix}
	  v_{1}^\alpha & i v_{3}^\alpha \\
          -i v_{3}^\alpha & v_{2}^\alpha
	  \end{pmatrix}, \\
             v_{1(2)}^\alpha &=& \frac{\partial
           \varepsilon_{1(2)}}{\partial  k_\alpha}, \quad
           v_{3}^\alpha =
	   \frac{\xi_{-}^2}{({\xi_{-}^2+\Delta_{\bf k}^2})^{1/2}}
	    \frac{\partial}{\partial k_\alpha} \frac{\Delta_{\bf k}}{\xi_{-}}.
            \label{velocities}
	    \end{eqnarray}
The off-diagonal term $v_3$ in the above
 expression arises from the $\bk$-dependence of the unitary
transformation $U$,  and corresponds to the interband transition
operator  $\Omega$. \cite{LL9}
The "mass" operator in the new basis is $U^\dagger \partial^\alpha
\partial^\beta{\hat H} U$. The explicit expression for it,
          \begin{equation}
          {\cal D}^\alpha {\cal D}^\beta \hat h =
          \frac {\partial^2 \hat h }
          {\partial  k_\alpha \partial  k_\beta}
          - \sigma^3
          \frac{2v_{3{\bf k}}^\alpha  v_{3{\bf k}}^\beta}
          {\varepsilon_{1{\bf k}}-\varepsilon_{2{\bf k}}},
          \label{mass}
          \end{equation}
is a smooth function in the whole Brillouin zone.

The scattering processes are modelled by the imaginary part $\gamma$
of the poles of Green's functions, so that ${\hat g}^{A(R)}_{11} =
(x- \varepsilon_1 \mp i \gamma)$.
In the limit of a large scattering time, $\tau = \gamma^{-1}$,  the
principal contribution to the conductivity (\ref{conduc2}) is delivered by
the combinations ${\hat g}^A_{11}(x) {\hat g}^R_{11}(x)$ and ${\hat
g}^A_{22}(x) {\hat g}^R_{22}(x)$,  where the poles of the Green's
functions differ only by the value for the damping.
One easily finds that this leading contribution to the
conductivity contains only "intraband" velocity terms. In the
limit of zero temperatures we have
      \begin{eqnarray}
      \sigma_{xx}  &=&
      e^2 \tau
      \int \frac{d\bk}
       {(2\pi)^2} \left [
       ({ v}_{1\bk}^x)^2
        \delta(\varepsilon_{1\bk}) + (1\leftrightarrow2)
      \right],
      \label{conduc3}
      \end{eqnarray}
in accordance with previous findings\cite{Chakravarty02,Norman03}.
Note that the "interband" current term $v_3$ in the
final expression for the conductivity is absent only in the d.c.\
limit, but is in general present in the optical
conductivity tensor and also modifies the optical sum rule. The optical sum is 
defined by the new mass (\ref{mass}), averaged over the occupied states
in the Brillouin zone, and should exhibit the deviations,  $\sim \Delta^2/E_F$ 
in the DDW state. 
   
%\section{Hall conductivity below transition temperature}
\label{sec:Hall2}

\begin{figure}
\includegraphics[width=8cm]{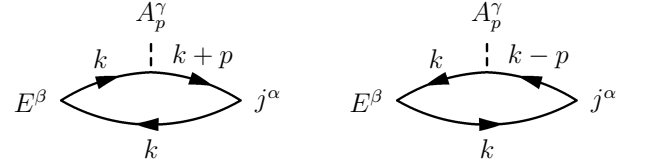}%
\caption{Two diagrams contributing to the
Hall conductivity
\label{fig:diagrams}}
\end{figure}

The next step is to evaluate the Hall conductivity tensor in the DDW
state. The magnetic field can be included by considering the first-order
change in the Green's functions due to the magnetic field in 
Eq.\ (\ref{conduc2}), as discussed in \cite{AKLL}.
Writing  ${\bf B}_\bp = i \bA_\bp \times \bp$ and taking eventually the
limit $p\to 0$, the change in the conductivity is described by the two diagrams
shown in Fig.\ \ref{fig:diagrams}.

The contribution from the left diagram in Fig.\  \ref{fig:diagrams}
assumes the form
        \begin{equation}
       e A^\gamma_\bp         {\rm Tr}\left[
       {\hat V}_{\bk}^\beta {\hat G}^A_{\bk}
       {\hat V}_{\bk+\bp/2}^\gamma {\hat G}^A_{\bk+\bp}
       {\hat V}_{\bk+\bp/2}^\alpha {\hat G}^R_{\bk}
       \right].
       \label{Hall1p}
      \end{equation}
The second diagram is obtained from the above expression by putting
 $\bp\to-\bp$
and applying hermitian conjugation corresponding to ${\hat G}^A
\to {\hat G}^R$.  The zeroth-order terms in $\bp$  in Eq.(\ref{Hall1p})
are odd in $\bk$ and vanish upon the subsequent integration over $\bk$.
The terms linear in $\bp$ assume the form $\frac{e} 2
A^\gamma_\bp p^\eta {\rm Tr} K^{\alpha\beta\gamma\eta}$ with the tensor
$K^{\alpha\beta\gamma\eta}$ given by
      \[
     {\hat G}^R_{\bk}
       {\hat V}_{\bk}^\beta {\hat G}^A_{\bk} \left[
       \frac{\partial
       {\hat V}_{\bk}^{\gamma}}{\partial k_\eta}
        {\hat G}^A_{\bk} {\hat
 V}_{\bk}^\alpha
       + 2 {\hat V}_{\bk}^\gamma
       \frac{\partial{\hat G}_{\bk}^A}{
       \partial k_\eta }      {\hat V}_{\bk}^\alpha
       + {\hat V}_{\bk}^\gamma {\hat G}^A_{\bk}
       \frac{\partial{\hat V}_{\bk}^{\alpha}}{\partial k_\eta}
\right] .   \]
Further steps include the use of the property
$\partial{\hat G}_{\bk}/\partial k_\eta =
{\hat G}_{\bk} {\hat V}_{\bk}^\eta {\hat G}_{\bk}$, the application
of the unitary transformation $U$ with the corresponding change
$\partial^\alpha \to {\cal D}^\alpha$, and an integration by parts over
$\bk$. Attention should be paid to the non-commutative property
of the involved matrices.  After some calculation 
$K^{\alpha\beta\gamma\eta}$ reduces to 
      \begin{equation}
      -{\cal D}^\eta( {\hat g}^R  {\hat v}^\beta )
        ({\cal D}^\gamma  {\hat g}^A) {\hat v}^\alpha
        + {\hat g}^R{\hat v}^\beta
	[{\hat g}^A{\hat v}^\gamma , {\hat g}^A{\hat v}^\eta]
       {\hat g}^A{\hat v}^\alpha.
      \end{equation}
Finally, we combine the expressions from the two diagrams in Fig.\
\ref{fig:diagrams} and retain the principal contribution in the
large-$\tau$ limit.  As a result we obtain for the
Hall current ${\bf j}$  in the low-temperature limit
        \begin{eqnarray}
	j^\alpha &=& \sigma_{\alpha\beta\zeta} E^\beta B^\zeta, \\
        \sigma_{\alpha\beta\zeta}  &=&
	e^3\tau^2 \epsilon_{\zeta\gamma \eta} \int \frac{d\bk}{(2\pi)^2}
	\left[ \delta(\ve_{1{\bf k}})
	v_{1{\bf k}}^\alpha v_{1{\bf k}}^\gamma \left(
	\frac{\partial^2 \varepsilon_{1{\bf k}}}
{\partial k_\beta\partial k_\eta}
	\right.\right.	\nonumber \\ &&\left.\left.
	+ \frac{2v_{3{\bf k}}^\beta  v_{3{\bf k}}^\eta}
        {\varepsilon_{1{\bf k}}-\varepsilon_{2{\bf k}}}
	\right)
         + (1\leftrightarrow 2)
	\right ],
	\label{main}
        \end{eqnarray}
with $ \epsilon_{\zeta\gamma \eta}$ being the totally antisymmetric tensor.

Eq.\ (\ref{main}) is the central result of this paper. It shows that the Hall
conductivity in the DDW state is defined by two inverse mass terms. The first
term ${\partial^2 \varepsilon_{1{\bf k}}}/
{\partial k_\beta\partial k_\eta}$ is the
direct analog of the standard expression \cite{Ziman} and is usually discussed
\cite{Chakravarty02,Norman03}. The second term 
${v_{3{\bf k}}^\beta  v_{3{\bf k}}^\eta}/(\varepsilon_{1{\bf k}}
-\varepsilon_{2{\bf k}})$ is also present 
in (\ref{mass}) but enters Eq.\ (\ref{main}) with an opposite sign. Let us 
discuss the relative importance of this term. 

First, this term contributes only in the anisotropic case and is unimportant 
particularly for an excitonic insulator \cite{HallExcIns}, which is described by 
the Hamiltonian (\ref{Ham-mat}) 
with $\xi_{\bf k} \propto {\bf k}^2$, $\xi_{{\bf k}+{\bf Q}}\propto -{\bf k}^2$ 
and $ \Delta_{\bf k} = constant$. 
In this case all three velocities,
${\bf v}_{1,2,3}$ in Eq.(\ref{velocities}), are
parallel to $\bk$. As a result, the second mass term in Eq.(\ref{main}), 
containing
${\bf v}_{1(2)} \times {\bf v}_{3}$, vanishes.

\begin{figure}
\includegraphics[width=8cm]{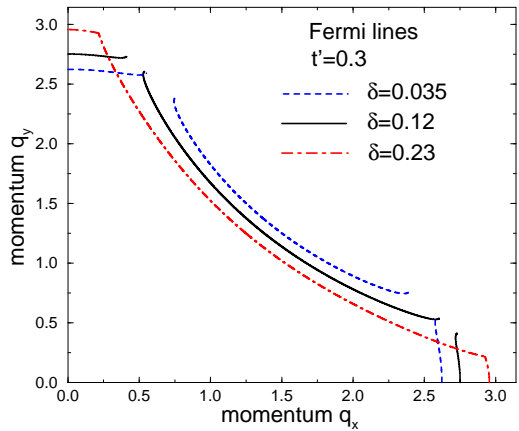}%
\caption{The evolution of the Fermi surface with the opening of the DDW gap. 
\label{fig:FerSur}}
\end{figure}

Second, quite generally, the interband 
current is present for an electron in a periodic potential, so that 
the analog of (\ref{main}) may occur in a multi-band metal as well. The 
main difference between this case and the discussed DDW state lies in 
the 
relative importance of the second inverse mass term in (\ref{main}). The 
energy denominator in it involves the interband splitting which is usually 
large in the multi-band case. The energies of two bands may 
become closer at the van Hove points in the Brillouin zone, however, the 
interband current $v_3$ vanishes there.
These general arguments are inapplicable to the DDW situation as described 
below.

For the anisotropic dispersion $\xi_\bk$ and DDW order parameter
$\Delta_\bk$ the second mass term in Eq.(\ref{main}) is important.
Indeed, the DDW-induced changes in the Hall conductivity are mostly determined
by the vicinity of the "hot spots" in $\bk-$space where
$\xi_{\bk}\simeq \xi_{\bk + \bQ}\simeq 0$. Expanding the spectrum around one
of these spots we write $\xi_{+} \simeq V_1 k_1$,  $\xi_{-} \simeq V_2
k_2$, and $\Delta_{\bk} \simeq \Delta_{hs}+V_d k_1$. Here  $k_{1,2} = k_x \pm
k_y $ and  $|\Delta_{hs}|\sim | V_d | \ll |V_1| \sim |V_2|$. 
These expressions lead to $v_3 \sim V_2$ and a parametrically small energy 
denominator $\varepsilon_{1{\bf k}}-\varepsilon_{2{\bf k}} \sim 
|\Delta|$ in (\ref{main}).
Eq.(\ref{main}) shows then an anomalously large contribution $\sim V_1^2 
V_2^2/|\Delta|$ in the hot spot's vicinity, $\delta k
\sim  |\Delta_0|/V_2$.  Observing that ${\cal D}^\alpha {\cal D}^\beta \hat h$ 
in (\ref{mass}) is finite near the hot spots, one expects that the second
mass term enhances substantially the anomalous contribution from the first term 
in (\ref{main}). 
The resulting change in the  Hall conductivity,  $\delta 
\sigma_{xyz}$, is estimated as       
       \begin{eqnarray}
       \delta\sigma_{xyz} &\simeq&
       e^3 \tau^2 2\pi^{-1} V_2 V_d sign(V_1 \Delta_{hs}).
       \label{sxy-corr}
       \end{eqnarray}
We see that  $\delta\sigma_{xyz}$ is negative for 
the above form of the spectrum, thus enhancing the absolute value of the 
(negative) $\sigma_{xyz}$.

We emphasize that the correction Eq.(\ref{sxy-corr}) which is linear in 
$|\Delta_0|$ 
explicitly contains the gap velocity $V_d$. In the case of an s-wave order 
parameter, 
$\Delta_{\bf k} = \Delta_0 = constant$, the velocity $V_d=0$ and the first 
nonvanishing correction to $\delta\sigma_{xyz}$ would be of order of $\Delta_0^2$ 
and thus much smaller. 

\begin{figure}
\includegraphics[width=8cm]{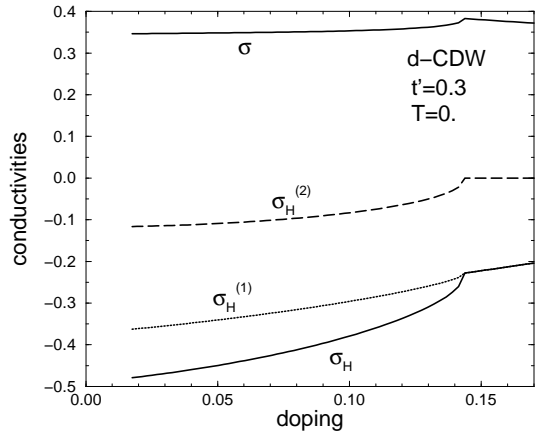}%
\caption{The zero-temperature results for the doping dependence of the 
conductivity $\sigma$ and the Hall conductivity $\sigma_H$ divided by $e^2 \tau 
$ and $e^3 \tau^2 $, respectively. The contributions 
from the first and second inverse mass term in (\ref{main}) to $\sigma_H$ 
are shown as $\sigma_H^{(1)}$ and $\sigma_H^{(2)}$, respectively. 
\label{fig:dopdep}}
\end{figure}

\begin{figure}
\includegraphics[width=8cm]{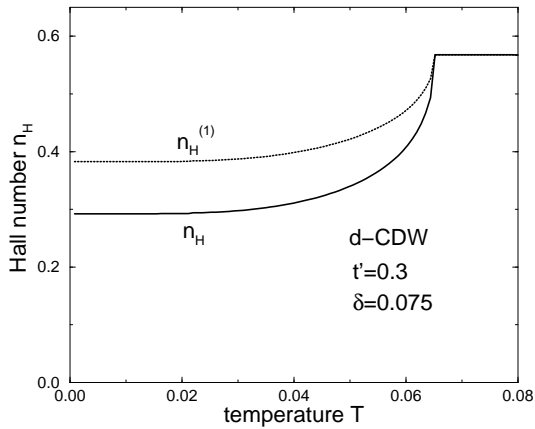}%
\caption{The temperature dependence of the Hall number 
$n_H = -\sigma^2/ \sigma_H$.  
\label{fig:temdep}}
\end{figure}

We have performed numerical calculations for $\sigma_{xx}=\sigma$
and $\sigma_{xyz}=\sigma_H$ using Eqs.(\ref{conduc3}) and (\ref{main}) 
and our Hamiltonian Eq.(\ref{H}). 
In rough agreement with the large-N limit of the $t-J$
model\cite{Cappelluti} we modelled the gap by 
$\Delta_0 = \bar{\Delta}(T)\sqrt(1+\mu)\Theta(1+\mu)$, where $\mu$ is the 
chemical potential, $\bar{\Delta}(0) = 0.58$, a BCS temperature
dependence is assumed for $\bar{\Delta}(T)$, and $t$ is used as the energy 
unit. The onset of the gap
at $\mu=-1$ corresponds to the critical doping $\delta_c \sim 0.145$ at $T=0$ and
to the critical temperature $T_c \sim 0.064$ at $\delta=0.075$, using always 
$t'=0.3$.
Fig.\ \ref{fig:FerSur} shows Fermi lines of this model for three different 
dopings. The Fermi lines consist of arcs around the nodal direction
and lines near the antinodal points. Lines for the same doping end at
the boundary of the magnetic Brillouin zone at different points because
of the presence of the gap.

The conductivities at zero temperature were obtained as integrals
over Fermi lines. We used several hundred points to parametrize
the Fermi lines ensuring that similar grids were used for
different lines to achieve a numerical cancellation of singular
terms. The temperature dependent conductivities
$\sigma(\mu,T)$ were calculated using
 \[
 \sigma(\mu, T) = \int dx \, \frac{\partial n (x)}{\partial
x}\sigma(\mu +x , 0) = \int_0^1 dn\,  \sigma(\mu +x(n) , 0), \]
with $x(n) = T \ln (n^{-1} -1 )$. The latter redefinition of the
integration regularizes the calculation at low temperatures.

The conductivity $\sigma$ has a contribution linear in the order parameter 
coming from the vicinity of hot spots,
$ \delta\sigma_{xx} \simeq - e^2 \tau \pi^{-1}| V_2 \Delta_{hs}/ V_1|$.
It translates to a square root dip near the
critical values $\delta_c$ and $T_c$, as can be seen 
in the curve for $\sigma$ in Fig.\ \ref{fig:dopdep} for the case of 
$\delta_c$. Assuming that most of the scattering is due to
impurities, $\tau$ is qualitatively unchanged at $T_c$. Consequently, the
square root feature should be observable not only in $\sigma_H$ but also
in $\sigma$. Note, however, that this dip in $\sigma_H$ and $\sigma$ 
is determined by $\frac{d\Delta_{\bf k}}{d{\bf k}}$ 
and $\Delta_{\bf k}$ at hot spots, 
respectively. 
As shown in Fig.\ \ref{fig:dopdep} the usual
expression for $\sigma_H$ (first term in Eq.(\ref{main}), denoted by
$\sigma_H^{(1)}$), exhibits only a
very weak change at $\delta_c$ as a function of doping. In contrast to
that, the new term (second term in Eq.(\ref{main}), denoted by
$\sigma_H^{(2)}$), shows a well-pronounced square-root
behavior near $\delta_c$ and dominates the change in the total Hall conductivity
$\sigma_H = \sigma_H^{(1)} + \sigma_H^{(2)}$. The temperature dependence of the 
conductivities is qualitatively similar to the doping one.  

Fig.\ \ref{fig:temdep} depicts the temperature dependence of the Hall number
$n_H=-{\sigma}^2/\sigma_H$. 
The curve denoted by $n_H^{(1)}$ is based on the usual expression, the
curve $n_H$ on our complete expression including the extra mass term.
The onset of the DDW again causes an approximate  square root decay below $T_c$
in both cases. From a quantitative point of view it is clear from this
Figure that the conventional theory gives only roughly 2/3 of the decay
so that the discovered new term cannot be neglected in quantiative calculations.

In conclusion, we derived an expression for the Hall conductivity $\sigma_H$
in the CDW state including also the interband current contribution.
As a result, there is an additional term to $\sigma_H$ which may be interpreted
as a renormalization of the mass and which is especially important for
momentum-dependent CDW order parameters. It is shown numerically that
the new term increases the anomalous contribution $\sim \sqrt{T_c-T}$ 
to $\sigma_H$ by about a factor 2 in the case of the DDW.

\end{document}